\documentclass[12pt]{myarticle}

\usepackage{txfonts}
\usepackage[colorlinks]{hyperref}

\newcommand{\n}{\noindent}
\newtheorem{Theorem}{Theorem}
\newtheorem{Proposition}{Proposition}
\newtheorem{lemma}{Lemma}
\newtheorem{cor}{Corollary}
\newtheorem{definition}{Definition}
\newtheorem{ex}{Example}

\begin{document}

\title{F-Index of Four Operations on Graphs}

\author{Nilanjan De\corref{cor1}}
\ead{de.nilanjan@rediffmail.com}

\address{Department of
Basic Sciences and Humanities (Mathematics),\\ Calcutta Institute of Engineering and Management, Kolkata, India.}
\cortext[cor1]{Corresponding Author.}

\begin{abstract}
The F-index of a graph is defined as the sum of cubes of the vertex degrees of the graph which was introduced in 1972, in the same paper where the first and second Zagreb indices were introduced. In this paper we study the F-index of four operations on graphs which were introduced by Eliasi and Taeri [M. Eliasi, B. Taeri, Four new sums of graphs and their Wiener indices, \textit{Discrete Appl. Math.}\textbf{157}(2009) 794--803.].

\end{abstract}
\maketitle

\section{Introduction}

Using chemical graph theory, different chemical structures are usually modeled by a molecular graph to understand different properties of the chemical compound theoretically. A molecular graph is a pictorial representation of the structural formula of a chemical structure where atoms of the given chemical compound are represented through vertices and the edges represent the chemical bonds between the atoms. Also a molecular graph must be an unweighted, undirected graph without self loop or multiple edges. A molecular topological index transforming chemical information of a molecular graph by means of a numeric parameter which characterize its topology and is necessarily invariant under automorphism of graphs. Topological indices correlates the physico-chemical properties of molecular graph and is nowadays a standard procedure in studying structure-property relations. In chemistry, biochemistry and nanotechnology topological indices have been found to be useful in isomer discrimination, quantitative structure-activity relationship (QSAR) and structure-property relationship (QSPR) for predicting different properties of chemical compounds and biological activities. Also, topological indices has shown its high applicability in the discovery and design of new drugs. Among different classes of topological indices, those based on vertex degree of the molecular graph are called vertex-degree based topological indices which are most studied and have good correlations to the chemical properties.

Let, $G=(V,E)$ be a connected, undirected simple graph with $|V|=n$ and $|E|=m$. Let $N(u)$ denotes the first neighbor set of $u$; then $|N(u)|={{d}_{G}}(u)$ is called the degree of the vertex $u$. As usual $P_n$ and $C_n$ denote a path and cycle graph of order $n$ respectively. The first and second Zagreb indices of a graph were introduced in 1972 \cite{gutm72}, denoted by $M_1(G)$ and $M_2(G)$ and are respectively defined as

\begin{center}
${{M}_{1}}(G)=\sum\limits_{v\in V(G)}{{{d}_{G}}{{(v)}^{2}}}=\sum\limits_{uv\in E(G)}{[{{d}_{G}}(u)+{{d}_{G}}(v)]}$~~ and~~  ${{M}_{2}}(G)=\sum\limits_{uv\in E(G)}{{{d}_{G}}(u){{d}_{G}}(v)}$.
\end{center}

These indices are among one of the most important vertex-degree based topological indices and are used to study molecular complexity, chirality, ZE-isomerism, and hetero–systems and hence attracted more and more attention from chemists and mathematicians (see \cite{xu15,das15,kha09,zho07,zho05}). In the same paper \cite{gutm72}, another topological index, defined as sum of cubes of degrees of the vertices of the graph was also introduced. Recently Furtula and Gutman in \cite{fur15} studied this index and establish some basic properties of this index and showed that the predictive ability of this index is almost similar to that of first Zagreb index and for the entropy and acetic factor, both of them yield correlation coefficients greater than 0.95. They named this index as  ``forgotten topological index" or "F-index". Very recently the present author studied this index for different graph operations \cite{de16a} and also introduced its coindex version in \cite{de16b}. De et al. in \cite{de16c} also studied F-index of different classes of nanostar dendrimers. In \cite{fur15a} Furtula et al. explore some basic properties and bounds of F-index and in \cite{abd15} Abdoa et al. investigate the trees extremal with respect to the F-index. Throughout this paper we named this index as F-index and denoted by $F(G)$, so that

\begin{eqnarray}
F(G)=\sum\limits_{v\in V(G)}{{{d}_{G}}{{(v)}^{3}}}=\sum\limits_{uv\in E(G)}{[{{d}_{G}}{{(u)}^{2}}+{{d}_{G}}{{(v)}^{2}}]}.
\end{eqnarray}

Li et al. in \cite{li05} introduced the general first Zagreb index of a graph and is defined as
\begin{eqnarray}
{{\xi }_{n}}(G)=\sum\limits_{v\in V(G)}{{{d}_{G}}{{(v)}^{n}}}=\sum\limits_{uv\in E(G)}{[{{d}_{G}}{{(u)}^{n-1}}+{{d}_{G}}{{(v)}^{n-1}}]}
\end{eqnarray}
where $n$ is an integer, not 0 or 1. For different study of these index see \cite{zha06,zha06a,gut14,li04}. Obviously ${{\xi }_{2}}(G)=M_1(G)$ and ${{\xi }_{3}}(G)=F(G)$.
One of the redefined versions of Zagreb index is defined as
\begin{eqnarray}
{Re}Z{{G}}(G)=\sum\limits_{uv\in E(G)}{{{d}_{G}}(u){{d}_{G}}(v)[{{d}_{G}}(u)+{{d}_{G}}(v)]}.
\end{eqnarray}
There are various recent study of redefined versions of Zagreb index, for details see \cite{ran13,ran16,gao16,xu12}.

Graph operations played a very important role in chemical graph theory. Barri\`{e}re et al. in 2009 \cite{bar09,barr09} generalized the concept of Cartesian product of graphs by introducing a new graph operation called generalized hierarchical product graph. Where as Eliasi et al. in \cite{eli09} introduced four new sum of graphs (called F-sums) and compute the Wiener index. However, it was found that, these four new sums of graphs can be considered as a special case of generalized hierarchical product of graphs. In this paper, we compute the F-index of F-sum graphs not from direct calculation but as an application of generalized hierarchical product of graphs and hence using the derived results we find F-index of some particular and chemically interesting graphs.

\section{Main Results}

Barri\`{e}re et al. in 2009 introduced a new graph operation called the hierarchical product of graphs \cite{bar09} and in the same year they also reported the generalized hierarchical product of graphs which is a generalization of both Cartesian product of graphs and hierarchical product of graphs \cite{barr09}. Several results on different topological indices under generalized hierarchical product of graphs are already studied \cite{are10,elia13,arez13,luo14,de14a}.

\begin{definition} Let $G$ and $H$ be two connected graphs and $U$ be a non-empty subset of $V(G)$. Then the hierarchical product of $G$ and $H$, denoted by $G(U)\Pi H$, is the graph with vertex set $V(G)\times V(H)$, and any two vertices $(u,v)$ and $({u}',{v}')$ of $G(U)\Pi H$ are adjoint by an edge if and only if $\Big[u={u}'\in U$ and $vv'\in E(H)\Big]$ or $\Big[v=v'$ and $uu'\in E(G)\Big]$.
\end{definition}

From definition, we can state the following lemma which gives some basic properties of generalized hierarchical product of graphs.

\begin{lemma}
(a) $G(U)\Pi H$ is connected if and only if $G$ and $H$ are connected.

(b) $|V(G(U)\Pi H)|=|V(G)||V(H)|$ and $|E(G(U)\Pi H)|=|E(G)||V(H)|+|E(H)||U|$.

(c) The degree of a vertex $(u_1,u_2)$ of ${{G}}(U)\Pi {{H}}$ is given by
\[{{d}_{G(U)\Pi H}}({u_i},{v_j}) = \left\{ \begin{array}{ll}
{{d}_{{{G}}}}(u_i)+{{d}_{{{H}}}}(v_j),{u_i}\in U\\[2mm]
{{d}_{{{G}}}}(u_i),{u_i}\in V({{G}})-U.
\end{array}\right.\]
\end{lemma}

First, we calculate the F-index of generalized hierarchical product of two graphs $G$ $(n\ge 2)$ and $H$. For details application of F-index of generalized hierarchical product of graphs see \cite{de16a}.

\begin{Theorem}
Let $G$ $(n\ge 2)$ and $H$ be two connected graphs. Then,
\begin{eqnarray}
F(G(U)\Pi H)=|V(H)|F(G)+6|E(H)|\sum\limits_{{{u}_{i}}\in U}{{{d}_{G}}{{({{u}_{i}})}^{2}}+}3{{M}_{1}}(H)\sum\limits_{{{u}_{i}}\in U}{{{d}_{G}}({{u}_{i}})}+|U|F(H)
\end{eqnarray}
\end{Theorem}
\n\textit{Proof.} From definition of generalized hierarchical product of graphs and F-index, we have

\begin{eqnarray*}
F(G(U)\Pi H)&=&\sum\limits_{i=1}^{|V(G)|}{\sum\limits_{j=1}^{|V(H)|}{{{\{{{d}_{G(U)\Pi H}}({{u}_{i}},{{v}_{j}})\}}^{3}}}}\\
            &=&\sum\limits_{{{u}_{i}}\in U}{\sum\limits_{j=1}^{|V(H)|}{{{\{{{d}_{G(U)\Pi H}}({{u}_{i}},{{v}_{j}})\}}^{3}}}}+\sum\limits_{{{u}_{i}}\in V(G)\backslash U}{\sum\limits_{j=1}^{|V(H)|}{{{\{{{d}_{G(U)\Pi H}}({{u}_{i}},{{v}_{j}})\}}^{3}}}}\\
            &=&\sum\limits_{{{u}_{i}}\in U}{\sum\limits_{j=1}^{|V(H)|}{{{\{{{d}_{G}}({{u}_{i}})+{{d}_{H}}({{v}_{j}})\}}^{3}}}}+\sum\limits_{{{u}_{i}}\in V(G)\backslash U}{\sum\limits_{j=1}^{|V(H)|}{{{d}_{G}}{{({{u}_{i}})}^{3}}}}\\
            &=&\sum\limits_{{{u}_{i}}\in U}{\sum\limits_{j=1}^{|V(H)|}{{{\{{{d}_{G}}{{({{u}_{i}})}^{3}}+{{d}_{H}}{{({{v}_{j}})}^{3}}+3{{d}_{G}}{{({{u}_{i}})}^{2}}{{d}_{H}}({{v}_{j}})+3{{d}_{G}}({{u}_{i}}){{d}_{H}}{{({{v}_{j}})}^{2}}\}}^{3}}}}\\
            &&+\sum\limits_{{{u}_{i}}\in V(G)\backslash U}{\sum\limits_{j=1}^{|V(H)|}{{{d}_{G}}{{({{u}_{i}})}^{3}}}}\\
            &=&|V(H)|\left[ \sum\limits_{{{u}_{i}}\in U}{{{d}_{G}}{{({{u}_{i}})}^{3}}}+\sum\limits_{{{u}_{i}}\in V(G)\backslash U}{{{d}_{G}}{{({{u}_{i}})}^{3}}} \right]+6|E(H)|\sum\limits_{{{u}_{i}}\in U}{{{d}_{G}}{{({{u}_{i}})}^{2}}}\\
            &&+3{{M}_{1}}(H)\sum\limits_{{{u}_{i}}\in U}{{{d}_{G}}({{u}_{i}})+}|U|F(H)\\
            &=&|V(H)|F(G)+6|E(H)|\sum\limits_{{{u}_{i}}\in U}{{{d}_{G}}{{({{u}_{i}})}^{2}}+}3{{M}_{1}}(H)\sum\limits_{{{u}_{i}}\in U}{{{d}_{G}}({{u}_{i}})+}|U|F(H).
\end{eqnarray*}
Hence the desired result follows.                                  \qed

\bigskip

The concept of F-sum graph was first introduced by \cite{eli09} and the Weiner indices of the resulting graphs were studied therein. Li et al. in \cite{li11} derived explicit expression of the PI indices of four sums of two graphs. The hyper and reverse Weiner indices of F-sum graphs were studied by Metsidik et al. in \cite{met10}. Eskender et al. \cite{esk13} determined the eccentric connectivity index of F-sum graphs in terms of some invariants of the factors. An et al. in \cite{ana14} derived two upper bounds for the degree distances of four sums of two graphs. Deng et al. in \cite{deng16} study the first and second Zagreb indices of this four operations on graphs. Let, $L(G)$ denotes the line graph of $G$ which is the graph with vertex set $E(G)$ and two vertices of $L(G)$ are adjacent if and only if the corresponding edges have a vertex in common. First we recall some relevant definitions an notations.

\begin{definition}

(a) The subdivision graph of a graph $G$, denoted by $S(G)$, is obtained from $G$ by replacing each edge of $G$ by a path of length two.

(b) The triangle parallel graph of a graph $G$ is denoted by $R(G)$ and is obtained from $G$ by replacing each edge of $G$ by a triangle.

(c) The line superposition graph $Q(G)$ of a graph $G$ is obtained from $G$ by inserting a new vertex into each edge of $G$ and then joining with edges each pair of new vertices on adjacent edges of $G$.

(d) The total graph $T(G)$ of a graph $G$  has its vertices as the edges and vertices of $G$ and adjacency in $T(G)$ is defined by the adjacency or incidence of the corresponding elements of $G$.
\end{definition}
For mathematical properties and applications of the these subdivision graphs, we refer the reader to \cite{nd15,nd15a,yan07}.

\begin{definition} Let F be one of the subdivision operations $S, Q, R$ or $T$. For two connected graphs $G$ and $H$, the F-sum, denoted by $G{{+}_{F}}H$, is the graph with vertex set $(V(G)\cup E(G))\times V(H)$, and any two vertices $(u,v)$ and $({u}',{v}')$ of $G{{+}_{F}}H$ are adjacent if and only if \Big[$u={u}'\in V(G)$ and $v{v}'\in E(H)$\Big] or \Big[$v={v}'\in V(H)$ and $u{u}'\in E(F(G))$\Big].
\end{definition}

The following lemma gives some basic properties of F-sum graphs.

\begin{lemma}
If $U=V(G)$, then we have\\
(i) $|V(S(G)|=|V(R(G))|=|V(Q(G))|=|V(T(G))|=|V(G)|+|E(G)|.$\\
(ii) $|E(S(G))|=2|E(G)|$

     $|E(R(G))|=3|E(G)|$

     $|E(Q(G))|=2|E(G)|+|E(L(G))|$

     $|E(T(G))|=3|E(G)|+|E(L(G))|.$\\
(iii) Let, $F=\{S,R,Q,T\}$, then for every vertex $v\in U$, we have

      ${{d}_{S(G)}}(v)={{d}_{Q(G)}}(v)={{d}_{G}}(v)$ and ${{d}_{R(G)}}(v)={{d}_{T(G)}}(v)=2{{d}_{G}}(v)$.\\
(iv) Also, for every vertex $v\in V(F(G))\backslash U$, we have

      ${{d}_{S(G)}}(v)={{d}_{R(G)}}(v)=2$and ${{d}_{Q(G)}}(v)={{d}_{T(G)}}(v)={{d}_{L(G)}}(v)+2.$
\end{lemma}

Now if we consider $U=V(G)\subseteq V(F(G)$, then $F(G)(U)\Pi H = G{{+}_{F}}H$. Thus, using Theorem 1, we can easily compute the F-index of these four graph operations. To do that, in the following, first we determine the F-index of different subdivision graphs of the vertices of $U$ in terms of other graph invariants (using equations (1), (2) and (3)). 

\begin{Proposition}
If $U$ be a non-empty subset, then

(i) $F(S(G)(U))=F(G)+8|E(G)|$

(ii) $F(R(G)(U))=8F(G)+8|E(G)|$

(iii) $F(Q(G)(U))=F(G)+{{\xi }_{4}}(G)+3{Re}ZM(G)$

(iv) $F(T(G)(U))=8F(G)+{{\xi }_{4}}(G)+3{Re}ZM(G).$
\end{Proposition}

\n\textit{Proof.}
(i) If $U$ be a non-empty subset of $V(S(G))$, we have by Lemma 2
\begin{eqnarray*}
F(S(G)(U))&=&\sum\limits_{v\in V(S(G)(U))}{{{d}_{S(G)(U)}}{{(v)}^{3}}}\\
          &=&\sum\limits_{v\in U}{{{d}_{S(G)}}{{(v)}^{3}}}+\sum\limits_{v\in V(S(G))\backslash U}{{{d}_{S(G)}}{{(v)}^{3}}}\\
          &=&\sum\limits_{v\in V(G)}{{{d}_{G}}{{(v)}^{3}}}+\sum\limits_{v\in V(S(G))\backslash U}{{{2}^{3}}}\\
          &=&F(G)+8|E(G)|.
\end{eqnarray*}
(ii) If $U$ be a non-empty subset of $V(R(G))$, we have by Lemma 2
\begin{eqnarray*}
F(R(G)(U))&=&\sum\limits_{v\in V(R(G)(U))}{{{d}_{R(G)(U)}}{{(v)}^{3}}}\\
          &=&\sum\limits_{v\in U}{{{d}_{R(G)}}{{(v)}^{3}}}+\sum\limits_{v\in V(R(G))\backslash U}{{{d}_{R(G)}}{{(v)}^{3}}}\\
          &=&\sum\limits_{v\in V(G)}{{{\{2{{d}_{G}}(v)\}}^{3}}}+\sum\limits_{v\in V(R(G))\backslash U}{{{2}^{3}}}\\
          &=&8F(G)+8|E(G)|.
\end{eqnarray*}

(iii) If $U$ be a non-empty subset of $V(Q(G))$, we have by Lemma 2
\begin{eqnarray*}
F(Q(G)(U))&=&\sum\limits_{v\in V(Q(G)(U))}{{{d}_{Q(G)(U)}}{{(v)}^{3}}}\\
           &=&\sum\limits_{v\in U}{{{d}_{Q(G)}}{{(v)}^{3}}}+\sum\limits_{v\in V(Q(G))\backslash U}{{{d}_{Q(G)}}{{(v)}^{3}}}\\
           &=&\sum\limits_{v\in V(G)}{{{d}_{G}}{{(v)}^{3}}}+\sum\limits_{v\in V(R(G))\backslash U}{{{\{{{d}_{L(G)}}(v)+2\}}^{3}}}\\
           &=&F(G)+\sum\limits_{uv\in E(G)}{{{[{{d}_{G}}(u)+{{d}_{G}}(v)]}^{3}}}\\
           &=&F(G)+{{\xi }_{4}}(G)+3{Re}ZM(G).
\end{eqnarray*}

(iv) If $U$ be a non-empty subset of $V(T(G))$, we have by Lemma 2
\begin{eqnarray*}
F(T(G)(U))&=&\sum\limits_{v\in V(T(G)(U))}{{{d}_{T(G)(U)}}{{(v)}^{3}}}\\
          &=&\sum\limits_{v\in U}{{{d}_{T(G)}}{{(v)}^{3}}}+\sum\limits_{v\in V(T(G))\backslash U}{{{d}_{Q(G)}}{{(v)}^{3}}}\\
          &=&\sum\limits_{v\in V(G)}{{{\{2{{d}_{G}}(v)\}}^{3}}}+\sum\limits_{v\in V(T(G))\backslash U}{{{\{{{d}_{L(G)}}(v)+2\}}^{3}}}\\
          &=&8F(G)+\sum\limits_{uv\in E(G)}{{{[{{d}_{G}}(u)+{{d}_{G}}(v)]}^{3}}}\\
          &=&8F(G)+{{\xi }_{4}}(G)+3{Re}ZM(G).
\end{eqnarray*}
Hence the desired result follows.            \qed

\begin{Theorem}
Let $G$ $(n\ge 2)$ and $H$ be two connected graphs. Then,
\[F(G{{+}_{S}}H)=|V(H)|F(G)+|V(G)|F(H)+6|E(H)|{{M}_{1}}(G)+6|E(G)|{{M}_{1}}(H)+8|V(H)||E(G)|.\]
\end{Theorem}

\n\textit{Proof.} From (4), using Proposition 1(i), we have
\begin{eqnarray*}
F(G{{+}_{S}}H)&=&F(S(G)(U)\Pi H)\\
              &=&|V(H)|F(S(G))+6|E(H)|\sum\limits_{{{u}_{i}}\in V(G)}{{{d}_{S(G)}}{{({{u}_{i}})}^{2}}+}3{{M}_{1}}(H)\sum\limits_{{{u}_{i}}\in V(G)}{{{d}_{S(G)}}({{u}_{i}})}\\
              &&+|V(G)|F(H)\\
              &=&|V(H)|\{F(G)+8|E(G)|\}+6|E(H)|\sum\limits_{{{u}_{i}}\in V(G)}{{{d}_{G}}{{({{u}_{i}})}^{2}}}+3{{M}_{1}}(H)\sum\limits_{{{u}_{i}}\in V(G)}{{{d}_{G}}({{u}_{i}})}\\
              &&+|V(G)|F(H)\\
              &=&|V(H)|F(G)+6|E(H)|{{M}_{1}}(G)+6|E(G)|{{M}_{1}}(H)+|V(G)|F(H)+8|V(H)||E(G)|.
\end{eqnarray*}
Hence the desired result follows.                    \qed

\begin{Theorem}
Let $G$ $(n\ge 2)$ and $H$ be two connected graphs. Then,
\[F(G{{+}_{R}}H)=8|V(H)|F(G)+|V(G)|F(H)+24|E(H)|{{M}_{1}}(G)+12|E(G)|{{M}_{1}}(H)+8|V(H)||E(G)|.\]

\end{Theorem}
\n\textit{Proof.} Applying Proposition 1(ii) in equation (4), we get
\begin{eqnarray*}
F(G{{+}_{R}}H)&=&F(R(G)(U)\Pi H)\\
              &=&|V(H)|F(R(G)(U))+6|E(H)|\sum\limits_{{{u}_{i}}\in V(G)}{{{d}_{R(G)}}{{({{u}_{i}})}^{2}}+}3{{M}_{1}}(H)\sum\limits_{{{u}_{i}}\in V(G)}{{{d}_{R(G)}}({{u}_{i}})}\\
              &&+|V(G)|F(H)\\
              &=&|V(H)|\{8F(G)+8|E(G)|\}+6|E(H)|\sum\limits_{{{u}_{i}}\in V(G)}{{{\{2{{d}_{G}}({{u}_{i}})\}}^{2}}}\\
              &&+3{{M}_{1}}(H)\sum\limits_{{{u}_{i}}\in V(G)}{2{{d}_{G}}({{u}_{i}})+}|V(G)|F(H)\\
              &=&8|V(H)|F(G)+8|V(H)||E(G)|+24|E(H)|{{M}_{1}}(G)\\
              &&+12|E(G)|{{M}_{1}}(H)+|V(G)|F(H).
\end{eqnarray*}
Hence the result. \qed

\begin{Theorem}
Let $G$ $(n\ge 2)$ and $H$ be two connected graphs. Then,
\begin{eqnarray*}
F(G{{+}_{Q}}H)&=&|V(H)|F(G)+|V(G)|F(H)+6|E(H)|{{M}_{1}}(G)+6|E(G)|{{M}_{1}}(H)\\
               &&+|V(H)|{{\xi }_{4}}(G)+3|V(H)|{Re}ZM(G).
\end{eqnarray*}
\end{Theorem}
\n\textit{Proof.} Using Proposition 1(iii) in equation (4), we get
\begin{eqnarray*}
F(G{{+}_{Q}}H)&=&F(Q(G)(U)\Pi H)\\
              &=&|V(H)|F(Q(G)(U))+6|E(H)|\sum\limits_{{{u}_{i}}\in V(G)}{{{d}_{Q(G)}}{{({{u}_{i}})}^{2}}+}3{{M}_{1}}(H)\sum\limits_{{{u}_{i}}\in V(G)}{{{d}_{Q(G)}}({{u}_{i}})}\\
              &&+|V(G)|F(H)\\
              &=&|V(H)|\{F(G)+{{\xi }_{4}}(G)+3{Re}ZM(G)\}+6|E(H)|\sum\limits_{{{u}_{i}}\in V(G)}{{{\{{{d}_{G}}({{u}_{i}})\}}^{2}}}\\
              &&+3{{M}_{1}}(H)\sum\limits_{{{u}_{i}}\in V(G)}{{{d}_{G}}({{u}_{i}})+}|V(G)|F(H)\\
              &=&|V(H)|F(G)+|V(H)|{{\xi }_{4}}(G)+3|V(H)|{Re}ZM(G)+6|E(H)|{{M}_{1}}(G)\\
              &&+6|E(G)|{{M}_{1}}(H)+|V(G)|F(H).
\end{eqnarray*}
Hence the desired result follows.                                 \qed

\begin{Theorem}
Let $G$ $(n\ge 2)$ and $H$ be two connected graphs. Then,
\begin{eqnarray*}
F(G{{+}_{T}}H)&=&8|V(H)|F(G)+|V(G)|F(H)+24|E(H)|{{M}_{1}}(G)+12|E(G)|{{M}_{1}}(H)\\
              &&+|V(H)|{{\xi }_{4}}(G)+3|V(H)|{Re}ZM(G).
\end{eqnarray*}
\end{Theorem}
\n\textit{Proof.} Applying Proposition 1(iv) in equation (4), we have
\begin{eqnarray*}
F(G{{+}_{T}}H)&=&F(T(G)(U)\Pi H)\\
               &=&|V(H)|F(T(G)(U))+6|E(H)|\sum\limits_{{{u}_{i}}\in V(G)}{{{d}_{T(G)}}{{({{u}_{i}})}^{2}}+}3{{M}_{1}}(H)\sum\limits_{{{u}_{i}}\in V(G)}{{{d}_{T(G)}}({{u}_{i}})}\\
               &&+|V(G)|F(H)\\
               &=&|V(H)|\{8F(G)+{{\xi }_{4}}(G)+3{Re}ZM(G)\}+6|E(H)|\sum\limits_{{{u}_{i}}\in V(G)}{{{\{2{{d}_{G}}({{u}_{i}})\}}^{2}}}\\
               &&+3{{M}_{1}}(H)\sum\limits_{{{u}_{i}}\in V(G)}{2{{d}_{G}}({{u}_{i}})+}|V(G)|F(H)\\
               &=&8|V(H)|F(G)+|V(H)|{{\xi }_{4}}(G)+3|V(H)|{Re}ZM(G)+24|E(H)|{{M}_{1}}(G)\\
               &&+12|E(G)|{{M}_{1}}(H)+|V(G)|F(H).
\end{eqnarray*}
Hence we get the desired result.  \qed

\bigskip

Let $U=V({{C}_{n}})$, where ${{C}_{n}}(n\ge 3)$ be a cycle, then from Theorem 2, we have the following corollary.

\begin{cor}
 Let $H$ be an arbitrary connected graph, then
\begin{equation}
F({C_n}{+_{S}}H) = nF(H)+6n{M_1}(H)+24n|E(H)|+16n|V(H)|.
\end{equation}
\end{cor}

\begin{ex} Let $\Gamma$ be the zigzag polyhex nanotube ${TUHC_6}[2n,2]$ with $n\ge 3$. Since $\Gamma \cong {{C}_{n}}{{+}_{S}}{{P}_{2}}$, using Corollary 1, we have
$F(TUH{{C}_{4}})=F({{C}_{n}}{{+}_{S}}{{P}_{2}})=70n.$
\end{ex}

\begin{cor}
 Let $H$ be an arbitrary connected graph, then
\begin{equation}
F({P_n}{+_{S}}H) = nF(H)+6n{M_1}(H)+12n(2n-3)|E(H)|+2n(8n-11)|V(H)|.
\end{equation}
\end{cor}

\begin{ex}
Let ${{L}_{n}}\cong {{P}_{n+1}}{{+}_{S}}{{P}_{2}}$ be a linear hexagonal chain with $n\ge 2$. Then using Corollary 2, we have $F({{L}_{n}})=F({{P}_{n+1}}{{+}_{S}}{{P}_{2}})=70n-22$.
\end{ex}

\begin{ex} Applying Theorem 2 to 5 we get the following results

(i) $F({{P}_{n}}{{+}_{S}}{{P}_{m}})=72nm-74n-82m+72$

(ii) $F({{P}_{n}}{{+}_{R}}{{P}_{m}})=224nm-182n-312m+216$

(iii) $F({{P}_{n}}{{+}_{Q}}{{P}_{m}})=128nm-74n-212m+72$

(iv) $F({{P}_{n}}{{+}_{T}}{{P}_{m}})=280nm-182n-442m+216.$\\
\end{ex}

\section{Conclusion}

In this paper, we compute the `forgotten topological index' of the generalized hierarchical product of graphs and then using the obtained result we find the results for four new sum of graphs. As an application, we calculate F-index of the zigzag polyhex nanotube ${TUHC_6}[2n,2]$, linear hexagonal chain $L_n$ and other particular graphs.

\section*{Competing Interests}

The author declares that there is no conflict of interests regarding the publication of this paper.

\end{document}